\documentclass[aps,prd,amssymb,showpacs,floatfix,twocolumn,superscriptaddress,nofootinbib]{revtex4-2}
\usepackage{amsmath,amsfonts,graphics,epsfig}

\begin{document}
\title{Pion gravitational form factors  at large momentum transfer in the
instant-form relativistic impulse approximation approach}

\author{A.F.~Krutov}
\email{a$_$krutov@rambler.ru}
\affiliation{Samara State Technical University, 443100 Samara,
        Russia}
\affiliation{Samara Branch, P.N.~Lebedev Physical Institute of the Russian Academy of Sciences, 443011 Samara, Russia}
\author{V.E.~Troitsky}
\email{troitsky@theory.sinp.msu.ru}
\affiliation{D.V.~Skobeltsyn
	Institute of Nuclear Physics,\\
	M.V.~Lomonosov Moscow State University, Moscow 119991, Russia}
\date{October 22, 2023}

\begin{abstract}
We extend our relativistic theory of gravitational structure of composite
hadrons to obtain the pion gravitational form factors at large momentum
transfers. The approach was used in the case of intermediate region of the
variable in our preceding works [Phys.Rev.D{\bf 103}, 014029 (2021)] and
[Phys.Rev.D{\bf 106}, 054013 (2022)].
The calculation is
carried out in the framework of a relativistic composite-particle model
complemented by the special relativistic form of impulse approximation.
It is found that in the limit of massless and pointlike
quarks, the obtained asymptotic expansion coincides with the
predictions of perturbative QCD for gravitational pion form factors.
The principal contribution to the  asymptotics, coinciding with the
predictions of QCD, is given by the relativistic effect of spin rotation.
In particular, the asymptotics of the $D$ form factor is completely
determined by this kinematic effect. Several restrictions on the allowed
form of gravitational form factors of quarks are derived.
\end{abstract}

\maketitle

{\section{Introduction}
The theory of the gravitational
structure of hadrons is in the focus of investigation during last
decades. Its current status is reviewed in \cite{BuE23} and for earlier
reviews see, e.g., \cite{LeL14, Ter16, PoS18, BuE18, CoL20}. The basic
mathematical object in this  theory is the operator of the energy-momentum
tensor (EMT) of the particle (see, e.g., \cite{LoC17}), and, respectively,
its Lorentz-covariant decomposition in terms of the gravitational form
factors (GFFs). GFFs encode key information including the mass and spin of
a particle, the less well-known but equally fundamental $D$-term ($D$ stands
for the German word {\it Druck} meaning pressure), as well as
the information about distributions of energy, angular momentum, and
various mechanical properties such as, e.g., internal forces inside the
system.

Since the gravitational interaction is very weak, the
\textit{direct} measurement of GFFs  cannot be carried out in
experiments today, nor in the foreseeable future. However, as it
turns out, information about the EMT can be extracted
\textit{indirectly}
(see useful discussion in {\cite{BuE23oct}). At present, one obtains the
information about the GFF mainly from the hard-exclusive processes described
in terms of unpolarized generalized parton distribution (GPDs) or in terms
of generalized distribution amplitudes (GDAs)
\cite{BuE23, BuE23oct, PoS18, KuS18, LoP22}. Several indirect measurements
are underway and some are planned (see, e.g., \cite{Chr22, Geo22, Kha22,
Bur22, And21}). In these experiments, kinematic ranges are extended
and the scale of momentum transfers is significantly enlarged ($Q^2 = -t =
q^2,\,q-$ momentum transfer). In this connection, it is interesting to
study GFFs at  large momentum transfer, up to the asymptotic region
$Q^2\to\infty$.

Another interesting point is the possible comparison of our purely
relativistic model results with the results of \cite{Tan18, ToM21, ToM22}
where strict predictions of perturbative quantum chromodynamics (QCD)
are given for the asymptotic behavior of pion GFFs.


At $Q^2\to\infty$, QCD gives a trustable description of hadron physics.
Results obtained
from the first principles within the framework of this generally accepted
fundamental theory of strong interactions should be considered as some
additional constraint on the GFFs calculated in other approaches. In
particular, this is the case for various formulations of the quark
model {\cite{PaB07, SuD20, TaL22, ChM22, LoS22, KrT21, KrT22}.
It seems that an analog of the correspondence principle should be required,
i.e. in the theories of the gravitational structure operating with quark
and gluon degreess of freedom, there must be a limiting transition, giving
for the GFFs the behavior coinciding with the predictions of the
perturbative QCD.

Since the asymptotics of the GFFs of composite hadrons depends on the
asymptotic behavior of the GFFs of the constituent quarks, then the
correspondence condition imposes restrictions on quark form factors,
the determination of which is also a relevant task
(see, e.g., \cite{MoM22}).

To complete the motivation, it is important to mention two more
points of different generality of significance.
First, the calculation of the  mean-square
mechanical radius requires information about the behavior of the
form factor $D$ in the entire region of momentum transfer, including
asymptotic \cite{PoS18}. Second, our calculation can shed light on the
 position of the boundary of a perturbative regime for GFFs.

The presented paper is devoted to the calculation of the asymptotic
behavior of the pion GFF. In our calculations
we use a particular variant of the instant-form (IF) Dirac \cite{Dir49}
relativistic quantum mechanics (RQM)
(see also \cite{LeS78, KeP91, Coe92, KrT09, Pol23}),
extended for composite systems (see, e.g.,
\cite{KrT02, KrT03, KrP16, KrT05}).  The approach was
successfully used to describe the pion electromagnetic form factor.
Recently we have shown \cite{KrT21, KrT22} that the pion GFFs
can be derived in the same formalism using the same approximations and the
same model parameters, adding only one new parameter fixed by fitting the
slope at zero of the normalized to pion $D$-term form factor $D$ of pion.
The present paper, in fact, generalizes this approach to a larger
region of momentum transfer. We will frequently refer to the
results of these two articles.

The basic features of our approach are the following
(see, e.g., Ref. \cite{KrP16}).
\begin{enumerate}
\item
The main differences between our version and the conventional IF
RQM are, firstly, the construction of the matrix elements of local
operators, which are based on the analog of the Wigner-Eckart theorem
for the Poincar\'e group \cite{KrT05,Edm55}, and, secondly, the
interpretation of the corresponding reduced matrix elements, that is form
factors, as generalized functions.
\item
We include the interaction in the composite
system by adding the interaction operator to the operator of the mass of
the free constituent system by analogy with the conventional IF of RQM.
\item
Note that it is possible to include the interaction in our approach by the
solutions of the Muskhelishvili-Omn\`es-type equations \cite{TrS69}.
These solutions represent wave functions of constituent quarks.
\item
It is
important to notice that the approach we use differs from the IF
\textit{per se}; it is rather fruitfully complemented by a modified
impulse approximation, MIA, constructed by making use of a
dispersion-relation approach in terms of the reduced matrix elements. So,
the Lorentz-covariance condition and the current conservation law are
satisfied automatically. The difference between MIA and conventional IA is
detailed in our paper \cite{KrT02}.
\end{enumerate}

In the context of the present paper, the following point is worth noting.
An important advantage of the approach we use is matching with the QCD
predictions in the ultraviolet limit, when quark masses are switched off,
as expected at high energies \cite{KrT98, TrT13, KrT17}. The model
reproduces correctly not only the functional form of the QCD asymptotics,
but also the numerical coefficient. The analogous result holds also for
the kaon \cite{TrT21}.

Integral representations for electromagnetic and gravitational form factors
of composite systems in our approach are given by double integrals of a
special form, which are analogs of dispersion integrals over the
composite-system mass \cite{TrS69} (see also \cite{AnS87}). We have proved
relevant theorems and formulas for asymptotic expansion of such integrals
at $Q^2\to\infty$ in \cite{KrT08}.

In the present work we calculate the GFFs asymptotics using a modified
impulse approximation (MIA) that we formulated and successfully exploited
earlier. MIA is relativistic by construction {\it  a priori}. As we realize
our study in IF RQM, it is natural to name our approximation as instant
form relativistic impulse approximation as it was done by the authors of
\cite{BuE23}.

Now (using actually \cite{KrT08, KrT21, KrT22}) we calculate the
asymptotics of the pion form factors $A$ and $D$ at $Q^2\to\infty$.
Since the pion GFFs depend on the choice of the model two-quark wave
function only weakly \cite{KrT22}, we use the wave functions of the
harmonic oscillator (the Gaussian one) for simplicity. The pion GFFs
obtained in such way decrease exponentially in $Q$ with power--law
corrections in $1/Q$. It is shown that if in the asymptotic expansion we
go to massless point constituents, i.e. in the limit of zero mass and zero
mean-square mass radius of quarks, then our asymptotics for the form
factor $A$ coincides with that predicted by the perturbative QCD
\cite{Tan18, ToM21, ToM22}.  This result is obtained with the same
quark form factors with logarithmic decay as we used previously
for pion GFFs at finite $Q^2$ in the works \cite{KrT21, KrT22}, and on the
pion electroweak structure \cite{KrT02, KrT03, KrT01, KrT98, TrT13,
KrT17}.

However, if we want to
prioritize the principle of correspondence, and to obtain, for low quark
mass, a power-law asymptotics of pion form factor $D$, then it is necessary
to modify the gravitational form factor $D$ of quarks, namely to go from
logarithmic decreasing with increasing momentum transfer to a power-law
one. This modification leads to nonzero rms of the mechanical radius of
the quark, in contrast to the logarithmic dependence that gives a zero
value for this quantity. Thus, in the domain where perturbative QCD is
applicable, its predictions can be obtained as a limiting case of our
formulation of the relativistic composite-particle model. Note that the
correspondence principle is satisfied here  actually due to the importance
and universality of the kinematical relativistic effect of spin rotation
\cite{KrT99}.

The structure of the paper is as follows. Section \ref{sec: Sec 2} is
devoted to a brief description of the instant-form relativistic impulse
approximation and calculation within its framework of the pion
gravitational form factors. In Section \ref{sec: Sec 3} the
procedure of asymptotic expansion of the pion gravitational form
factors is described and the corresponding asymptotic formulas are given.
In Section \ref{sec: Sec 4} we discuss physical consequences of the
obtained asymptotic expansion, in particular, the role of the
relativistic spin  effect, limitations on the used gravitational form
factors of constituent quarks and the possibility of obtaining in our
approach the asymptotics of the pion GFF which coincides with QCD
prediction. Section \ref{sec: Sec 5} contains the main conclusions of the
work. The Appendices contain the formulas for the Clebsch-Gordan
coefficients of Poincar\'e group and for the so-called free two-particle
gravitational form factors used in the modified impulse approximation.

\section{Instant-form relativistic impulse approximation and the
calculation of the pion GFFs}
\label{sec: Sec 2}

In the present section, for convenience of the reader,  we remind the
main stages of our approach used for the calculation of GFFs in immediately
preceding papers \cite{KrT21, KrT22}.
The approach is based on the instant form relativistic
quantum mechanics with a fixed number of particles (IF RQM)
(see \cite{Dir49, LeS78, KeP91, Coe92, KrT09}).

The difference between RQM
and its nonrelativistic analog comes down
to the difference between the algebra of the Poincar\'e group and the
algebra of the Galilean group at their realizations on the set of dynamic
observables of a composite system. A special feature of the Poincar\'e
algebra in comparison with the algebra of the Galilean group is the fact
that at additive inclusion of the constituent-interaction operator into
the zero component of the total momentum (into the operator of the total
energy), to preserve the corresponding algebraic structure it is
necessary to make operators of some other observables interaction-dependent,
too.
Different ways of incorporating
interaction into algebra lead to various forms of relativistic (Dirac)
dynamics. Notice, that to preserve the algebra of the Galilean
group at additive inclusion of the interaction to the zero component of
the total $4$-momenta does not require modification of other group
generators, and this leads to the only non-relativistic dynamics -- the
dynamics of the Schr\"odinger equation. Relativistic dynamics can be
classified into so-called kinematic subgroups, i.e. subgroups of
observables independent of the interaction.
The IF RQM has as
kinematic subgroup the group of motions of three-dimensional Euclidean
space, i.e. rotations and translations. One might say that
RQM occupies an intermediate position between the local
quantum field theory and non-relativistic quantum mechanics. Thus, the
constituents of the composite system are assumed to lie on the mass
shell, and the wave function of interacting particles is defined as an
eigenfunction of the complete set, which in IF RQM consists of
the following operators:

\begin{equation}
        {\hat M}_I^2\;
(\hbox{or}\;\hat M_I)\;,\quad
	{\hat J}^2\;,\quad \hat J_3\;,\quad \hat {\vec P}\;,
        \label{complete}
\end{equation}
where $\hat M_I$ is the mass operator for the
system with interaction, ${\hat J}^2$ is the operator of the square of the
total angular
momentum, $\hat J_3$ is the operator of the projection of the total
angular momentum on the $z$ axis and $\hat {\vec P}$ is the operator of the
total momentum.

In the IF RQM, the operators ${\hat J}^2\;,\; \hat J_3\;,\; \hat {\vec P}$
coincide with corresponding operators for the composite system without
interaction, and only the term $\hat M_I^2\;(\hat M_I) $
is interaction dependent.

To solve the problem on eigenfunctions of the set (\ref{complete}) it is necessary to
choose a suitable basis in the Hilbert state space of the composite system
(see details in \cite{KrT21}).
In the case of a system of two constituent quarks one can use, first, the
basis of individual spins and momenta:

\begin{equation}
	|\,\vec p_1\,,m_1;\,\vec p_2\,,m_2\,\!\rangle =
	|\,\vec p_1\,,m_1\,\! \rangle \otimes
	|\, \vec p_2\,,m_2\, \rangle\;,
        \label{p1p2}
\end{equation}
where $\vec p_1,\,\,\vec p_2$ are the 3-momenta of particles,
$m_1,\,m_2$ are the projections of spins to the $z$ axis.

Second, it is possible to use the basis in which the
motion of the center of mass of two particles is separated:
\begin{equation}
	|\,\vec P,\;\sqrt {s},\;J,\;l,\;S,\;m_J\,\rangle\;,
        \label{Pk}
\end{equation}
where $P_\mu = (p_1 +p_2)_\mu$, $P^2_\mu = s$, $\sqrt {s}$
is the invariant mass of the system of two
particles, $l$ is the orbital momentum in the center-of-mass of the system
(c.m.s.), $\vec S\,^2=(\vec S_1 + \vec S_2)^2 = S(S+1)\;,\;S$ is the
total spin in c.m.s., $J$
is the total angular momentum, $m_J$ is the projection of the total
angular momentum.

The bases (\ref{p1p2}) and (\ref{Pk}) are linked
by the Clebsch-Gordan decomposition of a direct
product (\ref{p1p2}) of two irreducible representations of the
Poincar\'e
group into irreducible representations (\ref{Pk}) \cite{KrT21, KrT09}.
The formulas for the corresponding Clebsch-Gordan coefficients are
given in Appendix A.

In the basis (\ref{Pk}) only the operator $\hat M_I$ in the complete
set (\ref{complete}) is non-diagonal. So, the two-quark wave function
in pion in the basis (\ref{Pk}) has the following form:
\begin{equation}
	\langle\vec P\,,\,\sqrt {s}\,|\,\vec p_\pi\rangle =
	N_C\,\delta (\vec P\, - \vec p_\pi)\,\varphi(s)\;,
        \label{wfI}
\end{equation}
where $\vec p_\pi$ is the 3-momentum of the pion. Explicit  form of the
normalization constant $N_C$ is given in the paper \cite{KrT21} and is
not used here. In the notations of basis vectors (\ref{Pk}), the
quantum numbers of the pion are omitted.

The wave function of the intrinsic motion is the eigenfunction of the
operator $\hat M_I^2\;(\hat M_I)$ and in the case of two particles of
equal masses is (see, e.g., \cite{KrT02})
$$
	\varphi(s(k)) = \sqrt[4]{s}\,k\,u(k)\;,\quad s = 4(k^2 + M^2)\;,
	$$\
\begin{equation}
	\int\,u^2(k)\,k^2\,dk = 1\;,
        \label{phi}
\end{equation}
where $u(k)$ is a model quark-antiquark wave function of the pion and
$M$ is the mass of the constituents.

Let us now construct the pion EMT in the IF RQM.
using the general method
of the relativistically invariant parametrization
of matrix elements of local operators
\cite{ChS63}  (see also \cite{CoL20}). For convenience and to preserve the
generality of the results of \cite{KrT21},  here we are dealing with the
same notations. The relation between our GFF and generally accepted
form factors $A$ and $D$ \cite{PoS18, Pag66} is given below in Section
\ref{sec: Sec 3}. For the pion EMT matrix element we obtained in \cite{KrT21}:
$$
\langle \vec p_\pi\left|T^{(\pi)}_{\mu\nu}(0)\right|\vec p_\pi'\rangle =
\frac{1}{2}G^{(\pi)}_{10}(Q^2)K'_\mu K'_\nu -
$$
\begin{equation}
	- G^{(\pi)}_{60}(Q^2)\left[Q^2g_{\mu\nu} + K_\mu K_\nu\right]\;,
        \label{Tpi}
\end{equation}
where $G^{(\pi)}_{10}, G^{(\pi)}_{60}$ are the gravitational form factors of the pion,
$g_{\mu\nu}$ is the metric tensor and
$$
K_\mu = (p_\pi - p_\pi')_\mu\,,\quad K'_\mu = (p_\pi + p_\pi')_\mu \;,
$$
$$
Q^2 = - t = - K_\mu^2
$$
We present the decomposition of the l.h.s.\ of
(\ref{Tpi}) over the basis  (\ref{Pk})
as a superposition of the same tensors
as in the r.h.s.\ of (\ref{Tpi}), and so obtain the pion GFFs in the
following form of the functionals defined on two-quark wave
functions (\ref{wfI}), (\ref{phi}):
$$
G^{(\pi)}_{i0}(Q^2) =
\int\,d\sqrt{s}\,d\sqrt{s'}\,
\varphi(s)\tilde G_{i0}(s,Q^2,s')\varphi(s')\;,
$$
\begin{equation}
	i=1,6\;.
        \label{int ds=Gpi}
\end{equation}
Here $\tilde G_{i0}(s,Q^2,s'),\,i=1,6$ are the
Lorentz-invariant regular distributions.

To calculate the invariant distributions on the r.h.s. of
(\ref{int ds=Gpi}), we use a version of impulse approximation.
The generally accepted impulse approximation (IA) is formulated in
the language of operators which means that EMT of the composite system
is assumed to be equal to the sum of single-particle EMT of the components:
\begin{equation}
	T \approx \sum_{k}\,T^{(k)}\;.
        \label{IA}
\end{equation}
To construct the pion GFFs we use
a modified impulse approximation (MIA) that we first formulated earlier
(see, e.g., Refs. \cite{KrT02,KrT03} and the review \cite{KrT09}) In
contrast to the baseline impulse approximation, MIA is formulated in terms
of the reduced matrix elements, that is form factors, and not in terms of
the operators itself. So, in MIA there appears important objects -- the
free gravitational form factors describing the gravitational
characteristics of systems without interaction.

Consider the system of two free constituent quarks \cite{KrT21}.
Note that in the work \cite{HuS18} it was shown that the form factor $D$
is zero in the case of point-like free fermions. In contrast, our
constituent quarks have all properties of realistic particles with
internal structure that is described by a set of form factors including
form factor $D$.

Using the general method of parametrization of local-operators matrix
elements \cite{ChS63} we write the matrix element of the EMT for system
of non-interacting fermions with total quantum numbers of pion
$J=l=S=0$ in the following form \cite{KrT21}:
$$
\langle P,\sqrt{s}\left|T^{(0)}_{\mu\nu}(0)\right|P\,',\sqrt{s'}\rangle =
$$
$$
= \frac{1}{2}G^{(0)}_{10}(s,Q^2,s')A'_\mu A'_\nu +
$$
\begin{equation}
	- G^{(0)}_{60}(s,Q^2,s')\left[Q^2\,g_{\mu\nu\nu} + A_\mu A_\nu\right]\;,
        \label{T0}
\end{equation}
where $G^{(0)}_{i0}(s,Q^2,s'),\,i=1,6$ are free two-particle GFFs,
In l.h.s. zero discrete quantum numbers in state
vectors are ignored.
$$
A_\mu = \left(P - P'\right)_\mu\;,\quad A^2 = t = -Q^2\;,
$$
$$
A'_\mu = \frac{1}{Q^2}\left[(s - s' + Q^2)P_\mu + (s' - s +
Q^2)P'_\mu\right]\;.
$$

The corresponding expression, based on the Clebsch-Gordan  decomposition,
that connects the bases (\ref{p1p2}) and (\ref{Pk}), for the EMT of a
system of non-interacting fermions with total quantum numbers of pion
$J=l=S=0$ in terms of the one-particle EMTs  has the following form
\cite{KrT21}:

$$
\langle P,\sqrt{s}\left|T^{(0)}_{\mu\nu}(0)\right|P\,',\sqrt{s'}\rangle =
$$
$$
= \sum\int\frac{d\vec p_1}{2p_{10}}\frac{d\vec p_2}{2p_{20}}
\frac{d\vec p\,'_1}{2p'_{10}}\frac{d\vec p\,'_2}{2p'_{20}}
\langle P,\sqrt{s}\left|\right.\vec p_1,m_1;\vec p_2,m_2\rangle\times
$$
$$
\left[\langle \vec p_1,m_1\left|\right.\vec p\,'_1,m'_1\rangle
\langle \vec p_2,m_2\left|T^{(u)}_{\mu\nu}(0)\right|\vec p\,'_2,m'_2\rangle\right. +
$$
$$
+ \left.\langle\vec p_2,m_2\left|\right.\vec p\,'_2,m'_2\rangle
\langle\vec p_1,m_1\left|T^{(\bar d)}_{\mu\nu}(0)\right|\vec p\,'_1,m'_1\rangle\right]\times
$$
\begin{equation}
        \langle \vec p\,'_1,m'_1;\vec p\,'_2,m'_2\left|\right.
        P\,',\sqrt{s'}\rangle\;,
        \label{T0p1p2}
\end{equation}
where $\langle P,\sqrt{s}\left|\right.\vec p_1,m_1;\vec p_2,m_2\rangle$ is
the Clebsch-Gordan coefficient (see Appendix A), the sums are over the variables
$m_1,\,m_2,\;m'_1,\,m'_2$.

The method gives for the one-particle matrix elements in
r.h.s.\ of (\ref{T0p1p2}) the form:
$$
\langle\vec p,m\left|T^{(q)}_{\mu\nu}(0)\right|\vec p\,',m'\rangle =
\sum_{m'}\langle m\left|D_w^{1/2}(p,p')\right|m'\rangle\times
$$
$$
\langle m''\left|(1/2)g^{(q)}_{10}(Q^2)K\,'_\mu K'_\nu
+ \right.
$$
$$
+ ig^{(q)}_{40}(Q^2)\left[K'_\mu\,R_\nu + R_\mu\,K'_\nu\right] -
$$
\begin{equation}
        \left. - g^{(q)}_{60}(Q^2)\left[Q^2g_{\mu\nu} + K_\mu K_\nu\right]
	\right|m'\rangle \;,
        \label{Tq}
\end{equation}
$q = u,\,\bar d$, $D_w^j(p,\,p')$ is the transformation operator from the
small group, the matrix of three-dimensional rotation,
$g^{(u,\bar d)}_{i0},\,i=1,4,6$ are the constituent-quark GFFs, the relation
of which with conventional notations is given below in Sect. \ref{sec: Sec
3}.
$$
K_\mu = (p - p')_\mu\,,\quad K'_\mu = (p + p')_\mu \,,
$$
\begin{equation}
        R_\mu = \epsilon _{\mu\,\nu\,\lambda
		\,\rho}\,p^\nu\,p'\,^\lambda\,\Gamma^\rho (p')\;.
        \label{kk'RG}
\end{equation}
Here $\Gamma^\rho (p')$ is well known 4-vector of spin (see, e.g.,
{\cite{ChS63, KrT21, KrT09, KrT03}),
$\epsilon _{\mu\,\nu\,\lambda\,\rho}$ is the absolutely antisymmetric
pseudotensor of rank 4, $\epsilon_{0\,1\,2\,3}= -1$.

Using the coefficients of the Clebsch-Gordan decomposition from Appendix A,
we obtain free two-particle form factors in (\ref{T0}) in terms of
gravitational form factors of constituents (\ref{Tq}). The corresponding
formulas are given in Appendix B.

Now let us exploit MIA in dealing with the obtained
system of equations for the free two-particle form factors.
MIA consists in replacing the invariant distribution in r.h.s.\ of
(\ref{int ds=Gpi}) by free two-particle form factors from Eq.~(\ref{T0}).
The physical meaning of
MIA is equivalent to that of the universally accepted IA (\ref{IA}),
because the free two-particle form factors are given in terms of
one-particle currents.

Thus, we get the expressions for
the pion GFFs in MIA, which for convenience we present
in the form given in \cite{KrT22}:

$$
G^{(\pi)}_{10}(Q^2) =
\frac{1}{2}\left[g^{(u)}_{10}(Q^2)+g^{(\bar d)}_{10}(Q^2)\right]\,G^{(\pi)}_{110}(Q^2) +
$$
\begin{equation}
	+ \left[g^{(u)}_{40}(Q^2)+g^{(\bar d)}_{40}(Q^2)\right]\,G^{(\pi)}_{140}(Q^2)\;,
        \label{Gpi1}
\end{equation}
$$
G^{(\pi)}_{60}(Q^2) =
\frac{1}{2}\left[g^{(u)}_{10}(Q^2)+g^{(\bar d)}_{10}(Q^2)\right]\,G^{(\pi)}_{610}(Q^2) +
$$
$$
+ \left[g^{(u)}_{40}(Q^2)+g^{(\bar d)}_{40}(Q^2)\right]\,G^{(\pi)}_{640}(Q^2) +
$$
\begin{equation}
	+ \left[g^{(u)}_{60}(Q^2)+g^{(\bar d)}_{60}(Q^2)\right]\,G^{(\pi)}_{660}(Q^2)\;,
        \label{Gpi6}
\end{equation}
where $g^{(q)}_{i0}(Q^2)\,,\,q=u,\bar d\,,\,i=1,4,6$ are the GFFs of the
constituent quarks, also defined previously in (\ref{Tq}).

The form factors in the r.h.s. of the equations (\ref{Gpi1}),
(\ref{Gpi6}) are given now in terms of the integrals \cite{KrT21}:
\begin{equation}
	G^{(\pi)}_{ij0}(Q^2) =\int d\sqrt{s} d\sqrt{s'}
	\varphi(s) G^{(0)}_{ij0}(s\,,Q^2\,,s')\varphi(s')\;,
        \label{Gpi1610}
\end{equation}
where
$i=1,6$; at $i=1\;j=1,4$; at $i=6\;j=1,4,6;$ $G^{(0)}_{1i0}(s\,,Q^2\,,s'),$
$\,G^{(0)}_{6k0}(s\,,Q^2\,,s')$
are components of the free GFFs that desribe the system of
two free particles with total quantum numbers of pion, given in Appendix B,
$\varphi(s)$ is the pion wave function in the sense of RQM
(\ref{phi}), $s'\,,s$ are the invariant masses of the free two-particle system in the
initial and final states, respectively.

Recall that the form factors $G^{(0)}_{k0}(s,Q^2,s')$ from (\ref{Gpi1610})
describe gravitational
features of a system of two particles without interaction.
Free two-particle form factors are regular generalized functions
(distributions) given by the corresponding functionals, defined
on the space of test functions depending on the variables ($s,\,s'$).
The functionals, in turn, are functions of the variable $Q^2 =-t$, a square
of momentum transfer. This variable is to be considered as a parameter.

In the frameworks of MIA, the pion GFFs are functionals (\ref{Gpi1}) --
(\ref{Gpi1610}), generated by the free two-particle GFFs of (\ref{T0}) on
test functions which are the products of the two-quark wave functions, see
(\ref{Gpi1}) -- (\ref{Gpi1610}).

\section{ASYMPTOTIC EXPANSION OF PION GRAVITATIONAL FORM FACTORS}
\label{sec: Sec 3}

In this section we calculate pion GFFs at large momentum transfer using
MIA in the IF RQM, that is in the instant form relativistic impulse
approximation. The conventional pion GFFs are connected with the matrix
elements given above by the equations:
\begin{equation}
	A^{(\pi)}(Q^2) = G^{(\pi)}_{10}(Q^2)\;,\;
	D^{(\pi)}(Q^2) = -2\,G^{(\pi)}_{60}(Q^2)\;,
        \label{ADpiG16}
\end{equation}
where $A^{(\pi)}$ and $D^{(\pi)}$ are commonly used pion GFFs
(see, e.g., \cite{PoS18, Pag66}), and $G^{(\pi)}_{10}$,
$\,G^{(\pi)}_{60}$  are
given by the equalities
(\ref{Gpi1}), (\ref{Gpi6}), $t = (p_\pi - p_\pi ')^2 = -Q^2$, and $\;p_\pi
', p_\pi$ are the pion 4-momenta in the initial and the final states,
respectively.

To obtain the asymptotic expansion of the pion form factors
given by (\ref{Gpi1610}) at $Q^2\to\infty$,
we use, for the quark-antiquark model wave function (\ref{phi}),
the wave function of
the ground state of a harmonic oscillator,
 widely
used in composite quark models:
\begin{equation}
u(k) =
2\left(1/(\sqrt{\pi}\,b^3)\right)^{1/2}\exp\left(-\,k^2/(2\,b^2)\right)\;.
\label{wfHO}
\end{equation}
We use the parameter $b$ in (\ref{wfHO}) fixed previously
in the works {\cite{KrT01, Kru97} on the electroweak.
properties of the pion.
It is important to note, that the actual choice of the model wave function
is not crucial in our approach.

As can be seen from the formulas for the pion form factors (\ref{Gpi1610}),
the asymptotics of the pion GFFs depends on the behavior of quark GFFs at
$Q^2\to\infty$.
Our GFFs of  quarks,
$g^{(q)}_{i0}(Q^2)\,,\,q=u,\bar d\,,\,i=1,4,6$ of (\ref{Tq}), are related
to generally accepted GFFs of a particle of spin
$1/2$ (see, e.g., \cite{PoS18}) as follows \cite{KrT21}:
$$
g^{(q)}_{10}(Q^2) = \frac{1}{\sqrt{1+Q^2/4M^2}}\left[\left(1 +\frac{Q^2}{4M^2}\right)\right.
A^{(q)}(Q^2) -
$$
\begin{equation}
        - \left. 2\frac{Q^2}{4M^2}J^{(q)}(Q^2)\right]\;,
        \label{g10}
\end{equation}
\begin{equation}
	g^{(q)}_{40}(Q^2) = -\,\frac{1}{M^2}\frac{J^{(q)}(Q^2)}{\sqrt{1 + Q^2/4M^2}}\;,
        \label{g40}
\end{equation}
\begin{equation}
	g^{(q)}_{60}(Q^2) = -\,\frac{1}{2}\sqrt{1 + \frac{Q^2}{4M^2}}D^{(q)}(Q^2)\;,
        \label{g60}
\end{equation}
where $A^{(q)},\,J^{(q)},\,D^{(q)}$ are the conventional GFFs of
particles with spin $1/2$ \cite{PoS18, Pag66}. We assume that the GFFs of $u$-
and $\bar d$-quarks
are equal: $g^{(u)}_{i0}(Q^2) = g^{(\bar d)}_{i0}(Q^2)\;, i=1,4,6$.
However, for more generality, now we relax the condition of one and the
same dependence of both $D$ and $A$ form factors of quark on momentum
transfer (as it was assumed in \cite{KrT22}):
$$
A^{(q)}(Q^2)=f^A_q(Q^2)\;,\quad J^{(q)}(Q^2) = \frac{1}{2}f^J_q(Q^2)\;,
$$
\begin{equation}
	D^{(q)}(Q^2) = D_q\,f^D_q(Q^2)\;,
        \label{AJDfq}
\end{equation}
where $D_q$ is the quark $D$-term. The functions in r.h.s.
must ensure the standard static limits (see, e.g.,
\cite{PoS18}):
\begin{equation}
	A^{(q)}(0)=1\;,\quad J^{(q)}(0) = \frac{1}{2}\;,\quad D^{(q)}(0) = D_q\;.
        \label{Ag0}
\end{equation}
The choice of functions in the right-hand sides  of (\ref{AJDfq}) will be
discussed in more detail below.

Let us make one more remark about the actual use of MIA
for calculation of the form factor $D$ of the pion
(\ref{ADpiG16}), (\ref{Gpi1610}). In our work \cite{KrT22}, it was found
that in MIA the $D$-form factor has a singularity at $Q^2=0$. So, to make
the form factor regular at zero, we were forced to abandon MIA in its
pure form and use some of its minimal generalization. In the
present work we are dealing with larger momentum transfer and
therefore we do not need to go beyond MIA.

The pion GFFs (\ref{Gpi1}) -- (\ref{Gpi1610}), are expressed in  our
approach in terms of double integrals of some special kind. In particular,
the boundary of the integration region (see the cutoff function
$\vartheta(s,Q^2,s')$ in Appendix B) depends on the parameter of
expansion, $Q^2$. The theorems defining the form of asymptotic
expansions of such integrals, and the corresponding formulas
were derived in our paper \cite{KrT08}. For
the integrals in (\ref{Gpi1}) -- (\ref{Gpi1610}) with wave function
(\ref{wfHO}) we have
(see \cite{KrT08}, Eqs. (55) -- (58) at $l=l'=$0):
$$
G^{(\pi)}_{ij0}(Q^2)\sim
$$
$$
\sim A_{ij}\exp\left(-\frac{1}{2\,b^2}\left(\frac{M}{2}\sqrt{Q^2 + 4\,M^2} - M^2\right)\right)
$$
\begin{equation}
\times \left[g^{(u)}_{j0}(Q^2)+g^{(\bar d)}_{j0}(Q^2)\right]\sum_{k=0}^\infty\sum_{m=0}^\infty\,h^{km}_{ij0}\;,
        \label{Gpi}
\end{equation}
$$
h^{km}_{ij0} =
$$
$$
= \sum_{p=0}^{p_m}\frac{b^{2m+2k-2p}}{Q^{2k+3m-5p-1/2}}\frac{2^{2m+5k/2-7p+5/2}}{M^{3p-m-1/2}}C^{4p}_{2m}\frac{(4p)!}{p!m!}
$$
$$
\times\frac{\partial^{2m-4p}}{\partial t^{2m-4p}}\left[G^{(0)(k)}_{ij0}\left(t\,,Q^2\,,\phi(t)\right)\right.
$$
\begin{equation}
\times\left.\left.\frac{\sqrt{(s-4M^2)(s'-4M^2)}}{\sqrt[4]{ss'}}\right]\right|_{t=0}\;,
\label{h}
\end{equation}
\begin{equation}
G^{(0)(k)}_{ij0}(t\,,Q^2\,,t') = \frac{\partial^k}{\partial t'\,^k}G^{(0)}_{ij0}(t\,,Q^2\,,t')\;,
                        \label{Gm}
\end{equation}
where $p_m=m/2+[(-1)^m - 1]/4$, the constant $A_{11}=A_{61}= 1/2$, all other constants $A_{ij}=1$,
$C^{4p}_{2m}$ is binomial coefficient. The variables $t,t'$ are linked to
the integration variables in (\ref{Gpi1610}) by the  relations:
$$
s = \frac{Q^2}{\sqrt{2}}\left(t' + t\right) + 2\,M^2 + M\sqrt{Q^2 + 4\,M^2}\;,
$$
\begin{equation}
s' = \frac{Q^2}{\sqrt{2}}\left(t' - t\right) + 2\,M^2 + M\sqrt{Q^2 + 4\,M^2}\;.
\label{ts}
\end{equation}
The function $t'=\phi(t)$ defines in variables
$t,t'$ a part of the boundary of the integration region in (\ref{Gpi1610})
containing the point $(t,t') = (0,0)$:
$$
t' = \phi(t) =
$$
\begin{equation}
=\sqrt{1 + \frac{4M^2}{Q^2}}\left(-\frac{\sqrt{2}\,M}{Q} + \sqrt{\frac{2M^2}{Q^2} + t^2}\right)\;.
\label{phit}
\end{equation}
The neighborhood of this point gives the
main contribution to the asymptotic expansion of the integrals. The total
boundary of the region of
integration in (\ref{Gpi1610}) in variables $(s,s')$ is given by
the cutoff function $\vartheta(s,Q^2,s')$ given in Appendix B.

We derive the asymptotic expansion  of pion GFFs (\ref{ADpiG16})
by means of general formulas (\ref{Gpi}) -- (\ref{phit}), taking
into account the expressions (\ref{Gpi1}) -- (\ref{ADpiG16}), (\ref{g10})
-- (\ref{AJDfq}). Thus, in each term containing the functions
$f^i_q(Q^2),\;i=A,J,D$ (\ref{AJDfq}), we leave only two
main terms of the $1/Q$ expansion, but the order of  the retained
terms must not exceed $1/Q^2$.  So, the  form factor $A$ of the pion
at $Q^2\to\infty$ is:
$$
A^{(\pi)}(Q^2) = G^{(\pi)}_{10}(Q^2) \sim \exp\left(-\frac{MQ}{4\,b^2}\right)\exp\left(\frac{M^2}{2\,b^2}\right)
$$
$$
\times \left[-2\sqrt{2}\,f^A_q(Q^2)\left(-4\frac{M}{Q} + 20\frac{M^2}{Q^2} + 24\frac{b^2}{Q^2} + 2\frac{M^4}{b^2\,Q^2}\right)\right. -
$$
$$
-2\sqrt{2}\,\left(f^A_q(Q^2) - f^J_q(Q^2)\right)\frac{Q^2}{4M^2}
$$
$$
\times\left(-4\frac{M}{Q} + 20\frac{M^2}{Q^2} + 24\frac{b^2}{Q^2} + 2\frac{M^4}{b^2\,Q^2}\right) +
$$
\begin{equation}
+ \left.96\sqrt{2}\,f^J_q(Q^2)\frac{b^2}{Q^2}\right]\;.
\label{Aas}
\end{equation}
The analogous expansion of the form factor $D$ is of the form:
$$
D^{(\pi)}(Q^2) = -2\,G^{(\pi)}_{60}(Q^2) \sim \exp\left(-\frac{MQ}{4\,b^2}\right)\exp\left(\frac{M^2}{2\,b^2}\right)
$$
$$
\times \left[4\sqrt{2}\,D_q\sqrt{1 + \frac{Q^2}{4M^2}}\,f^D_q(Q^2)\left(1 - \frac{M}{Q} - \frac{M^3}{2b^2\,Q}\right)\right. -
$$
\begin{equation}
	- \left.32\sqrt{2}\,f^J_q(Q^2)\frac{b^2}{Q^2}\right]\;.
        \label{Das}
\end{equation}
In the form-factor $D$ expansion the term proportional to $f^A_q(Q^2)$
(see Eqs. (\ref{Gpi6}), (\ref{g10}), (\ref{AJDfq}))}
is of order $\sim 1/Q^4$ and therefore
is not included in (\ref{Das}).

The consequences of the Eqs. (\ref{Aas}), (\ref{Das}) will be discussed in
the next Section.

\section{DISCUSSION OF RESULTS}
\label{sec: Sec 4}

From the expansions (\ref{Aas}), (\ref{Das})  it can be seen that the
pion GFFs in our relativistic approach and for the Gaussian type
model wave functions  (\ref{wfHO}) show at
$Q^2\to\infty$  the exponential decay in the
parameter $Q$. It is interesting to compare this result with that of
the nonrelativistic case. Note, that in the nonrelativistic limit the
integrals (\ref{Gpi1610}) with $u$ from (\ref{wfHO}) can be calculated
analytically. So, the corresponding nonrelativistic asymptotics, for
example, of the form factor $A$ is of the form:
$$
A^{(\pi)}_{NR}(Q^2) = G^{(\pi)}_{10\,NR}(Q^2) \sim
$$
\begin{equation}
       \sim f^A_q(Q^2)\exp\left(-\frac{Q^2}{16\,b^2}\right)\;.
        \label{AasNR}
\end{equation}
So, in nonrelativistic limit of our approach we obtain the Gaussian
(\ref{AasNR})  decay in parameter $Q$ at large momentum transfer.
Thus, the exponential decay in Eqs.
(\ref{Aas}), (\ref{Das}) is a strictly relativistic effect. It is a
consequence of our essentially relativistic nonperturbative approach.

Decay in (\ref{Aas}),
(\ref{Das}) is rather fast and provides, in particular, a finite
value of the rms mechanical radius of pion  $\langle r^2\rangle_{mech}$ ,
defined as follows (see, e.g., \cite{PoS18}):
\begin{equation}
        \langle r^2\rangle_{mech} = 6\frac{D(0)}{\int_0^{\infty}\,D^(Q^2)dQ^2}\;,
        \label{rqmech}
\end{equation}
The exponential decay of pion form factor $D$ (\ref{Das})
ensures the convergence of the integral at the upper limit
in (\ref{rqmech}) and, thus, its finite value.
The obtained decreasing of GFFs  (\ref{Aas}), (\ref{Das}) differs
fundamentally from the power-law form in perturbative QCD
\cite{Tan18, ToM21, ToM22} $\sim 1/Q^2$. In this latter case the
integral in the denominator of (\ref{rqmech}) diverges, and the rms
mechanical radius is zero. However, it can be shown that in our approach
there exists a limiting transition to the results coinciding with the
predictions of QCD, i.e., a kind of correspondence principle is satisfied.
We formulate it as follows. At large momentum transfer, where perturbative
QCD is applicable, its predictions can be obtained as certain limiting
case of our nonperturbative asymptotics (\ref{Aas}), (\ref{Das})}.

In this connection, let us first discuss  the asymptotic expansion of the
form factor $A$ of the pion (\ref{Aas}).
To ensure a correct transition in the Eq.(\ref{Aas}) to small masses of
constituents $M\to 0$, we require the equality:
$f^A_q(Q^2)=f^J_q(Q^2)$. Then at $M\to 0$ the Eq.(\ref{Aas}) takes
the form:
$$
A^{(\pi)}(Q^2) = G^{(\pi)}_{10}(Q^2) \sim
$$
\begin{equation}
\sim -48\sqrt{2}\,f^A_q(Q^2)\frac{b^2}{Q^2} + 96\sqrt{2}\,f^J_q(Q^2)\frac{b^2}{Q^2}\;.
        \label{AasM0}
\end{equation}
In the expansion (\ref{AasM0}) there is no exponent, the presence of  which
in (\ref{Aas}) is due to the type of two-quark wave function
(\ref{wfHO}), i.e., to the elastic coupling between quarks. Thus, the
transition to zero mass in the asymptotic region $Q^2\to\infty$
effectively eliminates any manifestation of quark-antiquark interactions.
Note that the second term in the sum (\ref{AasM0}) has a
purely relativistic origin and is a consequence of the relativistic
effect of spin rotation \cite{KrT99}. At neglecting this
effect this summand vanishes (see (\ref{Gpi1610}) and Eq.
(B2) in Appendix B).

Let us now discuss the constraints on the functions $f^A_q(Q^2)$  and
$f^J_q(Q^2)$ which can be derived using the asymptotic expansion
(\ref{AasM0}). In our previous work \cite{KrT22} we calculated the
pion GFFs at finite momentum transfer assuming that these functions
are  equal to one another, that is to the same function $f_q(Q^2)$:
$$
f^A_q(Q^2) = f^J_q(Q^2) =
$$
\begin{equation}
= f_q(Q^2) = \frac{1}{1 + \ln\left(1 + \langle r^2_q\rangle Q^2/6\right)}\;,
        \label{fq}
\end{equation}
where $\langle r^2_q\rangle$ is the rms mass radius
of the constituent quark, which we choose to be
equal to its charge radius.

The function (\ref{fq}) has been used successfully in our works on the
electroweak properties of the pion to describe
electric and magnetic form factors of constituent quarks. This function
was derived from asymptotics of the charge form factor of the pion at
$Q^2\to\infty$ \cite{KrT01}. We note that at large momentum transfer
the function (\ref{fq}) decreases as the inverse of the logarithm of
$Q^2$. Because of this slow decreasing of quark form factors, such a
choice corresponds to quark which is close to the point quark.

The choice of the form (\ref{fq})
causes the appearence of logarithmic multipliers in the expansion
without changing the actual powers of $1/Q^2$ in (\ref{AasM0}):
\begin{equation}
	A^{(\pi)}(Q^2) = G^{(\pi)}_{10}(Q^2) \sim
	48\sqrt{2}\frac{b^2}{Q^2}f_q(Q^2)\;.
        \label{AMfq}
\end{equation}
In (\ref{AMfq}) we have for the form factor $A$ of
pion, with the acceptable accuracy, up to logarithmic
multiplier, the same power-law  decrease that is obtained in the
perturbative QCD \cite{ToM21}. To calculate GFFs at finite momentum
transfer \cite{KrT21, KrT22} we fixed the parameter $\langle
r^2_q\rangle$, entering (\ref{fq}), basing on the works \cite{VoL90,
PoH90, TrT94, CaG96}: $\langle r^2_q\rangle \simeq {0.3}/{M^2}$.  However,
this relation is valid only for finite masses of constituents in
hadrons. In our papers \cite{KrT21, KrT22, KrT01, KrT02, KrT03},  we have
fixed this parameter at the value of the mass of constituents $M=0.22$
GeV, which has long ago become generally accepted in relativistic
calculations (see, e.g., \cite{GoI85, HaA23}). In the present work in our
asymptotic calculations we consider the mean-square mass radius of the
constituents as a free parameter.

If we go to point quarks, i.e., when $\langle r^2_q\rangle =0$ in
(\ref{fq}), simultaneously with the limiting transition to the zero mass
of constituents, that is, in fact, from constituent
quarks to current quarks, we get the asymptotics, coinciding with the
predictions of the perturbative QCD \cite{ToM21}:
\begin{equation}
	A^{(\pi)}(Q^2) = G^{(\pi)}_{10}(Q^2) \sim
	48\sqrt{2}\frac{b^2}{Q^2}\;.
        \label{AM}
\end{equation}
Note that this result coincides, up to a numerical multiplier,  with the
result obtained in our approach for the asymptotics of the charge form
factor of the pion \cite{KrT01}. The correct (positive) sign
of (\ref{AM}) is due to the presence of the second term of the sum in
(\ref{AasM0}), i.e. to the relativistic effect of spin rotation. It is
interesting that the asymptotics of (\ref{AM}) in the pointlike limit for
quarks $\langle r^2_q\rangle\to 0$ takes place for any choice of
decreasing function in (\ref{fq}), provided that  this function has the
form of the product $\langle r^2_q\rangle Q^2$.

Let us consider now the asymptotic expansion of the pion $D$-form factor
and, in particular, discuss the constraints, which can be derived,
concerning the possible form of the quark form factor $D$ (\ref{AJDfq}).
If we take the function $f^D_q(Q^2)$ entering the $D$-form factor of the
quark (\ref{AJDfq}) in the form with logarithm (\ref{fq}), used  at finite
$Q^2$ \cite{KrT22}, then from the asymptotic expansion (\ref{Das})
at $Q^2\to\infty$ and in the limit of small $M$ we obtain an
increasing, i.e. unphysical behavior of the  form factor $D$ of the pion.
Thus, for this choice of the quark  form factor $D$
there is no limit transition from the constituent quarks to current
quarks. So, to satisfy the correspondence principle it is necessary to
choose some other form for this quark form factor.

There is one more argument in favor of changing the form
of function $f^D_q(Q^2)$ in (\ref{AJDfq}), (\ref{Das}), a rather simple
one.  Indeed, when we choose $f^D_q(Q^2)$ in the form
(\ref{fq}) the rms mechanical radius of the constituent quark
$\langle r_q^2\rangle_{mech}$ is zero,
as can be seen from the  formula analogous to (\ref{rqmech}): the integral
in the denominator diverges at upper limit.

On the other hand, we consider a
constituent quark as a quasiparticle, that has all the properties of a
real particle, in particular, a non-zero mean-square mass radius
$\langle r^2_q\rangle$ (\ref{fq}). To obtain a non-zero value also of the
rms mechanical radius of the quark it is
necessary to have for its form factor $D$ a function with power-law
decreasing in $1/Q^2$ with the power higher than one.
To satisfy the condition of decreasing of the pion form
factor $D$ at $Q^2\to\infty$ in the (\ref{Das}) in the limit of small quark
masses simultaneously with the condition of finiteness of the rms of the
mechanical radius of the constituent quark allows, in
particular, for the following choice of the function $f^D_q(Q^2)$:
\begin{equation}
	f^D_q(Q^2) = \frac{1}{\sqrt{1 + Q^2/4M^2}}\frac{1}{1 + \langle r^2_q\rangle Q^2/6}\;.
        \label{fDq}
\end{equation}
Note that the power-law dependence of the electromagnetic form factors of
constituent quarks, similar to (\ref{fDq}),
was considered in \cite{CaG96}.

Taking into account the explicit form of the function (\ref{fDq}),  the
main terms of the expansion (\ref{Das}) in the limit of
almost zero masses of constituents, we obtain:
$$
D^{(\pi)}(Q^2) = -2\,G^{(\pi)}_{60}(Q^2) \sim
$$
\begin{equation}
\sim 4\sqrt{2}\,\frac{D_q}{1 + \langle r^2_q\rangle Q^2/6} - 32\sqrt{2}\,f^J_q(Q^2)\frac{b^2}{Q^2}\;.
\label{DasfD}
\end{equation}
It can be seen that when $Q^2\to\infty$, the expression (\ref{DasfD}) means
that the form factor $D$ of the pion behaves as $\sim 1/Q^2$, coinciding
with the predictions of QCD. The second summand in (\ref{DasfD}), due to
the choice of the function $f^J_q(Q^2)$ in the form (\ref{fq}), contains a
logarithmically decreasing multiplier. The second term in the sum in
(\ref{DasfD}), as well as in (\ref{AasM0}) is due to the
relativistic effect of spin rotation (see formulas (\ref{Gpi1610}) and
(B4) in Appendix B). This term has a decisive meaning in both cases.

Let us now move on in our discussion of (\ref{DasfD}) to point  quarks,
i.e. to current quarks. There is a general result for the $D$-term of
the point-like non-interacting fermion with spin $1/2$ \cite{PoS18, HuS18},
namely, it was shown that the $D$-term of such a fermion is zero. If
appeal to the concept of asymptotic freedom at $Q^2\to\infty$, then
the transition from constituent to current quark means $D_q=0$ and
$\langle r^2_q\rangle =0$ in expressions (\ref{AJDfq}), (\ref{DasfD}).
Thus, when going to the point quarks, the expansion (\ref{DasfD})
takes the form:
\begin{equation}
D^{(\pi)}(Q^2) = -2\,G^{(\pi)}_{60}(Q^2) \sim - 32\sqrt{2}\,\frac{b^2}{Q^2}\;.
        \label{DM}
\end{equation}
The resulting formula for the asymtotics of the form factor $D$ of  the
pion coincides with predicted by QCD and differs by a
numerical multiplier from the asymptotics of the $A$-form factor
(\ref{AM}), as takes place in  QCD as well \cite{ToM21}. So, a kind of
correspondence principle is valid for the pion form factor $D$, too. Note
that analogous correspondence was obtained in our works on electroweak
properties of scalar mesons, where we described the electromagnetic form
factors of pion and kaon at large momentum transfer
\cite{KrT17, TrT21}.
In the limit of zero-mass point quarks, our approach again appears to
be common to problems of electroweak and gravitational structures of the
pion. In the electroweak case the coincidence with perturbative
QCD predictions was obtained not only in calculations with Gaussian
functions of harmonic oscillator (\ref{wfHO}), but also with other
functions \cite{KrT98}, in particular with rational functions
\cite{CoP05}. We expect the same for the gravitational case.

The asymptotics (\ref{DM}), which coincides with the QCD
predictions, in the limit of the point quarks
($D_q=0,\;\langle r^2_q\rangle =0$), can be obtained for rather arbitrary
choice of second multiplier in the quark form factor $D$ (\ref{fDq}),
for example, any power of the multiplier can be used instead.

As can be seen from (\ref{AM}), (\ref{DM}), the multiplier before the
asymptotics depends on the parameter of our model $b$, which determines
the actual scale of the confinement. In calculations of the
pion GFFs at finite momentum transfer \cite{KrT21, KrT22} in the model
(\ref{wfHO}), as well as in our previous successful works on
the electroweak properties of the pion \cite{KrT01, KrT02, KrT03}, we have
used value $b=0.35$ GeV. This value gives good results also in
other forms of RQM, for example, in calculation of electroweak decays of
pions in the point-form  dynamics \cite{HaA23}.

Note  that the asymptotics of the form factor $D$ of the pion  in the
limit of point quarks remains to be completely determined by the second
summand in (\ref{DasfD}), i.e.\ completely is due to the relativistic
effect of spin rotation. Thus, in the asymptotic expansions (\ref{AM}),
(\ref{DM}), obtained in our approach and coinciding with the predictions
of the perturbative QCD, it is the kinematical relativistic effect of spin
rotation that plays a determining role.

Let us discuss another consequence of choosing the function $f^D_q(Q^2)$
in the form (\ref{fDq}). In this case we obtain the rms mechanical
radius of the quark (\ref{rqmech}) in the following form:
$$
\langle r_q^2\rangle_{mech} =
\frac{\sqrt{6\langle r^2_q\rangle}}{4M}\sqrt{1 - 2M^2\langle
r^2_q\rangle/3}
$$
\begin{equation}
\times\frac{1}{\arccos\left(\sqrt{2M^2\langle
r^2_q\rangle/3}\right)}\;.
\label{rqmechex}
\end{equation}
It can be seen from (\ref{rqmechex}), that the mechanical radius of
the constituent quark is zero in the limit of zero mass radius. This
expression also can be used in the calculation of gravitational properties
of pions and other composite hadrons at finite momentum transfer and finite
masses of constituents.

\section{Conclusions}
\label{sec: Sec 5}

In this paper, the asymptotic expansions of $A$ and $D$ pion GFFs
are obtained at large momentum transfer, $Q^2\to\infty$. We used a
version of the instant-form of Dirac relativistic quantum mechanics with
fixed number of particles (IF RQM)  complemented by an essentially
relativistic variant of impulse approximation, formulated previously
in our papers on the electroweak structure of composite particles. The
calculation was carried out in the model of interaction of constituent
quarks with quadratic confinement, namely with two-quark wave functions of
the ground state of the harmonic oscillator, i.e., with Gaussians. The
asymptotic decreasing of the form factors obtained in the paper is
exponential in the parameter $Q$, multiplied by a polynomial in $1/Q$.
In the non-relativistic case the analog of our model approach with
Gaussian wave function gives the Gaussian decreasing of the GFFs for
increasing $Q$. So, the asymptotic exponential decrease of the GFFs in
this interaction model is a consequence of our fundamentally essentially
relativistic study of the problem.

It is shown that for the obtained asymptotic relativistic expansion there
exists a limit transition from constituent to current quarks, i.e.,
the transition to point-like quarks of almost zero mass. This limit
obtained in our principally nonperturbative approach, gives the asymptotics
of gravitational form factors, coinciding with the predictions of the
perturbative QCD ($\sim 1/Q^2$) \cite{Tan18, ToM21, ToM22}. Thus,
an analog of the correspondence principle is satisfied in the following
terms: in the region where the perturbative QCD is applicable, its
predictions can be obtained as a limit case of the fundamentally
nonperturbative relativistic model of composite paeticles. Note that this
correspondence principle is fulfilled also in the case of electromagnetic
structure of the pion and kaon.

From the coincidence principle we obtained certain constraints on the
allowed form of quarks GFFs and
proposed simple formulas for them. Such kind of constraints  are  currently
in demand (see, for example, \cite{MoM22}) because of the existing
arbitrariness in calculations of hadron form factors in composite models.
To obtain the quark GFFs we use not only constraints on the piom GFFs
asymptotics, but also the property of the finiteness of the rms mechanical
radius of quark.

An equation is also derived connecting the rms mechanical and mass
radii of constituent quarks, (\ref{rqmechex}),
which can be used in the calculations of
gravitational properties of composite systems.

In this work we find that the determining contribution to the asymptotics
of pion GFFs, in the limit of massless and point-like quarks, coinciding
with the predictions of the perturbative QCD, comes in our approach
from the relativistic kinematical effect of rotation of quark spins (the
Wigner spin rotation). In particular, the correct, i.e., coinciding with
predictions of the QCD asymptotics of the  pion form factor $D$ is
completely determined by this kinematic effect.

To summarize, in this paper we have increased the scope of our research of
our previous papers \cite{KrT21, KrT22}
to the range of large momentum transfer and, by the way, obtained several
results that can be used elsewhere.

\section*{Appendix A. The Clebsch-Gordan coefficient}
The Clebsch-Gordan coefficient of the Poincar\'e group (\ref{T0p1p2}):

$$
\langle\,\vec p_1\,,m_1;\,\vec p_2\,,m_2\,|\,\vec P,\;\sqrt {s},
\;J,\;l,\;S,\;m_J\,\rangle =
$$
$$
=\sqrt {2s}[\lambda
(s,\,M^2,\,M^2)]^{-1/2}\,2P_0\,\delta (P - p_1 - p_2)
$$
$$
\times \sum \langle\,m_1|\,D_w^{1/2}(p_1\,,P)\,|\tilde m_1\,\rangle
\langle\,m_2|\,D_w^{1/2}(p_2\,,P)\,|\tilde m_2\,\rangle
$$
$$
\times \langle{1}/{2}\,{1}/{2}\,\tilde m_1\,\tilde
m_2\,|S\,m_S\,\rangle \,Y_{lm_l}(\vartheta\,,\varphi )
$$
$$
\times\langle S\,l\,m_S\,m_l\,|Jm_J\rangle\;,
\eqno{(A1)}
$$
where $\vec p = (\vec p_1 - \vec p_2)/2$, $p = |\vec p|$,
$\vartheta \,,\varphi$ are the spherical angles of the vector $\vec p$ in
c.m.s., $Y_{lm_l}$ is the spherical function,
$\langle\,S\,m_S\,|1/2\,1/2\,\tilde m_1\,\tilde m_2\,\rangle $ and
$\langle Jm_J|S\,l\,m_S\,m_l\,\rangle $ are the Clebsh-Gordan coefficients
of the group $SU(2)$, $\langle\,\tilde m|\,D_w^{1/2}(P,p)\,|m\,\rangle $
is the matrix of the
three-dimensional spin rotation, that is necessary for the
relativistic  invariant summation of the particle spins,
the sum being over $\tilde m_1,\,\tilde m_2,\,m_l,\,m_S$,
$$
\lambda(a,b,c) = a^2 + b^2 +c^2 - 2(ab + ac + bc)\;,
\eqno{(A2)}
$$
$M$ is the mass of constituent quarks.

We use in present paper the Clebsch-Gordan coefficient with pion quantum numbers
$J=l=S=0$.

\section*{Appendix B. Free two-particle gravitational form factors}

GFF of two noninteracting fermion with spin 1/2 (\ref{Gpi1610}):
$$
G^{(0)}_{110}(s, Q^2, s') = -\,\frac{R(s, Q^2, s')\,Q^2}{\lambda(s,-Q^2,s')}
$$
$$
\times\left[(4M^2+Q^2)\lambda(s,-Q^2,s') - \right.
$$
$$
- \left.3\,Q^2(s + s'+ Q^2)^2\right]\cos(\omega_1+\omega_2)
\;,
\eqno{(B1)}
$$
$$
G^{(0)}_{140}(s, Q^2, s') = -3\,M\,\frac{R(s, Q^2, s')\,Q^4}{\lambda(s,-Q^2,s')}
$$
$$
\times\xi(s,Q^2,s')(s+s'+Q^2)\sin(\omega_1+\omega_2)\;,
\eqno{(B2)}
$$
$$
G^{(0)}_{610}(s, Q^2, s') = \frac{1}{2}\,R(s, Q^2, s')
$$
$$
\times\left[(s + s' + Q^2)^2 - \right.
$$
$$
- \left.(4M^2 + Q^2)\lambda(s,-Q^2,s')/Q^2\right]\cos(\omega_1+\omega_2)
\;,
\eqno{(B3)}
$$
$$
G^{(0)}_{640}(s, Q^2, s') = -\,\frac{M}{2}\,R(s, Q^2, s')
$$
$$
\times\xi(s,Q^2,s')(s+s'+Q^2)\sin(\omega_1+\omega_2)\;,
\eqno{(B4)}
$$
$$
G^{(0)}_{660}(s, Q^2, s') = R(s, Q^2, s')
$$
$$
\times\lambda(s,-Q^2,s')\cos(\omega_1+\omega_2)\;,
\eqno{(B5)}
$$
where
$$
R(s, Q^2, s') = \frac{(s + s'+Q^2)}{2\sqrt{(s-4M^2) (s'-4M^2)}}\,
$$
$$
\times\frac{\vartheta(s,Q^2,s')}{{[\lambda(s,-Q^2,s')]}^{3/2}}\;,
$$
$$
\xi(s,Q^2,s')=\sqrt{-(M^2\lambda(s,-Q^2,s')-ss'Q^2)}\;,
$$
$\omega_1$ and $\omega_2$ are the Wigner spin-rotation parameters:
$$
\omega_1 =
\arctan\frac{\xi(s,Q^2,s')}{M\left[(\sqrt{s}+\sqrt{s'})^2 + Q^2\right] + \sqrt{ss'}(\sqrt{s} +\sqrt{s'})}\;,
$$
$$
\omega_2 = \arctan\frac{ \alpha (s,s') \xi(s,Q^2,s')} {M(s + s' + Q^2) \alpha (s,s') + \sqrt{ss'}(4M^2 + Q^2)}\;,
$$
$\alpha (s,s') = 2M + \sqrt{s} + \sqrt{s'} $,
$\vartheta(s,Q^2,s')= \theta(s'-s_1)-\theta(s'-s_2)$,
$\theta$ is the Heaviside function.
$$
s_{1,2}=2M^2+\frac{1}{2M^2} (2M^2+Q^2)(s-2M^2)
$$
$$
\mp \frac{1}{2M^2} \sqrt{Q^2(4M^2+Q^2)s(s-4M^2)}\;,
$$
$\lambda(a,b,c)$ and $M$ are determined in (A2).

\end{document}